\documentstyle[11pt]{article}
\title{The Open Path Phase for Degenerate and Non-degenerate
 Systems and its Relation to the Wave-function Modulus}
\author{R. Englman$^{1,2}$, A. Yahalom$^3$ and M. Baer$^1$\\
$^1$Department of Physics and Applied Mathematics\\
Soreq NRC, Yavne 81800, Israel\\
$^2$Research Institute,
College of Judea and Samaria, Ariel 44284, Israel\\
$^3$Faculty of Engineering,
Tel-Aviv University, Tel-Aviv 69978, Israel\\
(e-mail: englman@vms.huji.ac.il)}
\begin{document}
\maketitle

\newcommand{\beq}{\begin{equation}}
\newcommand{\enq}{\end{equation}}
\newcommand{\ber}{\begin{eqnarray}}
\newcommand{\enr}{\end{eqnarray}}
\newcommand{\nsc}{necessary and sufficient conditions }
\newcommand{\nec}{necessary }
\newcommand{\su}{sufficient }
\newcommand{\dl}{dimensionless }
\newcommand{\st}{stability }
\newcommand{\sa}{stationary }
\newcommand{\eq}{equation }
\newcommand{\eqs}{equations }
\newcommand{\br}{barotropic }
\newcommand{\er}[1]{\eq (\ref{#1}) }
\newcommand{\dr}[1]{definition (\ref{#1}) }

\begin{abstract}

We calculate the open path phase in a two state model with a slowly (nearly adiabatically)
varying time-periodic Hamiltonian and trace its continuous development during a period.
We show that the topological (Berry) phase attains $\pi$ or $2 \pi$
 depending on whether there is or
is not a degeneracy in the part of the parameter space enclosed by the trajectory.
Oscillations are found in the phase. As adiabaticity is approached, these become both more frequent and less pronounced and the phase jump becomes increasingly more steep.
Integral relations between the phase and the amplitude modulus (having the form of
Kramers-Kronig relations, but in the time domain) are used as an alternative way to
calculate open path phases. These relations attest to the observable nature of the
open path phase.

\bigskip


\end{abstract}

\section{Introduction}
\label{intro}

In the last fifteen years much attention has been given to phases in wave functions and,
in particular, to the topological (or Berry) phase, which is a signature of the trajectory of
the system (\cite{Berry}-\cite{Moore}) and which is manifest in some interference
and other experiments \cite{Bitter}.
As noted in earlier works \cite{Berry,EB,Resta},
the topological phase $\pm \pi$ that is picked up in a full
revolution of the system is linked to the existence of a degeneracy of states (or crossing
of potential energy surfaces) somewhere in the parameter space. This degeneracy need not be
located in a region that is accessed to in the revolution; however, its removal even by a
minute amount will cause the topological phase to be zero or an integral multiple of $2 \pi$.
The physical model treated in this work confirms this effect; indeed the calculated
topological phase shows a change from $\pi$ to $2\pi$ as the degeneracy disappears
(cf. Figs \ref{1} and \ref{2}).
 We tackle the problem by tracing a continuous variation in the non-cyclic,
open path phase \cite{Pati} (also named "connection" \cite{Simon}),
that is denoted in this work by
$\gamma (t)$ ($t$ is time).

To obtain an expression for $\gamma (t)$ we study (for both the degenerate and non-degenerate
 alternatives) an explicitly solvable model. Both a detailed analysis and the figures
exhibit, as a novelty, oscillations in $\gamma (t)$. These become increasingly more frequent
and of lesser magnitude as the adiabatic limit is approached. Furthermore it is
observed that in the adiabatic limit of the degenerate case, the change in the open
path phase is abrupt and results in a step function like behavior.

In an alternative approach to the calculation of the open-path phase we develop
reciprocal relations between phase and amplitude moduli of time dependent
wave functions (Section \ref{THY}). Versions of these relations in other contexts were
given earlier (\cite{Wolf}-\cite{Baer}).
The existence of these relations has the remarkable
consequence that the associated open path phase, defined by them, is a
"physical observable" (and {\it inter alia} gauge invariant) as a function of the path,
a quality heretofore associated with the closed path (Berry) phase.

\section{Theory}
\label{THY}

We start by invoking the Cauchy's integral formula which takes the form:
\beq
w(z)= \frac{1}{2 \pi i}\oint \frac{w(\zeta)}{\zeta - z} d \zeta
\label{Cauchy}
\enq
where $w(z)$ is analytic in the region surrounded by the anti-clockwise closed path.
In what follows we choose the closed path to be
the real axis t traversed in the reverse direction,
of the infinite interval
 $ -\infty \leq t \leq \infty $ and an infinite semi circle in the lower
half of the complex plane, as will be discussed later.
We shall concentrate on the case that $z$ is a real
variable $t$ and so \er{Cauchy} becomes:
\beq
w(t)= -\frac{1}{\pi i} P \int_{-\infty}^{\infty} \frac{w(t')}{t'-t} d t'
+ \frac{1}{\pi i} \oint_{SC} \frac{w(\zeta)}{\zeta - t} d \zeta
\label{Cauchyr}
\enq
where $P$ stands for the principal value of the integral,
$\zeta = \tau \exp{i \theta}, d \zeta = i \tau \exp{i \theta} d\theta$ and
it is assumed that $\tau \rightarrow \infty$ (the subscript $SC$ in the second term stands
for semi-circle).
Next it is assumed that $w(z)$ along the semi-circle is zero namely
\beq
\lim_{z \rightarrow \infty} w(z) = 0, \qquad {\rm for} \qquad \theta \neq 0,\pi
\label{conver}
\enq
so that \er{Cauchyr} becomes:
\beq
w(t)= -\frac{1}{\pi i} P \int_{-\infty}^{\infty} \frac{w(t')}{t'-t} d t'.
\label{Cauchyrc}
\enq
Assuming that the function $w(z)$ is written as $w(z) = w(t,y) =
u(t,y)+iv(t,y)$ where $z = t+iy$,  it can be shown by separating the real and the imaginary
parts, that \er{Cauchyrc} yields the two equations:
\beq
u(t)= - \frac{1}{\pi} P \int_{-\infty}^{\infty} \frac{v(t')}{t'-t} d t'
\qquad  {\rm and} \qquad
v(t)= \frac{1}{\pi} P \int_{-\infty}^{\infty} \frac{u(t')}{t'-t} d t'
\label{Cauchyrc2}
\enq
These relations are of the Kramers-Kronig (KK) or dispersion equations type
 \cite{Nussenzweig} and they
will be applied in the time domain. $u$ and $v$ are Hilbert transforms \cite{Titchmarsh}.
Our aim is to employ \eqs (\ref{Cauchyrc2}) to form a relation
between the phase factor in a wave
function and its amplitude-modulus. If a wave-function amplitude $ \tilde \psi(t)$
is written in the form:
\beq
\tilde \psi(t)= \tilde \Gamma(t) \exp (i \lambda (t))
\label{Gamlam1}
\enq
where $\Gamma(t)$ and $\lambda (t)$  are real functions of a real variable $t$,
the function $w(z)$ which will be defined as:
\beq
w(z) = \ln(\tilde \psi(z))= \ln(\tilde \Gamma(z))+ i \lambda (z)
\label{Gamlam2}
\enq
is assumed to fulfill the necessary conditions to employ the KK equations.
 This implies
the following: (a) the function $ \tilde \psi(z)$ is analytic and is free of zeroes
 in the lower complex half plane (however, $ \tilde \psi(z)$
 can have simple zeros on the real axis,
as is made clear in several publications
\cite{Wolf, Shapiro, Titchmarsh, Rev}).
(b) $ \tilde \psi(z)$ becomes, along the corresponding infinite
semi-circle, a constant (in fact this constant has to be equal to $1$ but if the constant is
$ \neq 1$
the analysis will be applied to $\tilde \psi(z)$ divided by this constant).
Thus, identifying $\ln(\tilde \Gamma(t))$  with $u(t)$ and $\lambda (t)$ with $v(t)$ we get
from the second part in \er{Cauchyrc2} the following expression:
\beq
\lambda (t)= \frac{1}{\pi} P \int_{-\infty}^{\infty}
\frac{\ln(\tilde \Gamma(t'))}{t'-t} d t'
\label{lamkk}
\enq
Next assuming that $\Gamma(t')$ is an even function,
the equation for $\lambda (t)$ can be written as:
\beq
\lambda (t)= \frac{2 t}{\pi} P \int_{0}^{\infty}
\frac{\ln(\tilde \Gamma(t'))}{{t'}^2-t^2} d t'
\label{lamkk2}
\enq
and if $\Gamma(t')$ is periodic then \er{lamkk2} can be further simplified to become:
\beq
\lambda (t)= \frac{2 t}{\pi} P \int_{0}^{\tilde T} d t' \ln(\tilde \Gamma(t'))
\sum_{n=0} \frac{1}{(t'+ N \tilde T)^2-t^2}
\label{lamkk3}
\enq
where $\tilde T$ is the relevant period.

\section{The Model}
\label{TM}

\subsection{The Basic Equations}
\label{TBE}

In this work, the reciprocal relations in \er{Cauchyrc2} are used in the form shown in
\er{lamkk3}.
A more general formulation of the reciprocal relations, including several applications,
will be presented in a separate publication.
Equation (\ref{lamkk3}) is applied to two examples based on the Jahn-Teller model \cite{Englman}
which, following Longuet-Higgins \cite{Longuet}, can be expressed in terms of an extended
version of the Mathieu equation, (\cite{EB,Charutz,Yahalom}) namely:
\beq
H = -\frac{1}{2} E_{el} \frac{\partial^2}{\partial \theta^2} - G_1 (q,\phi)\cos (2 \theta)
+ G_2 (q,\phi)\sin( 2 \theta).
\label{ham}
\enq
Here $\theta$ is an angular (periodic) electronic coordinate, $\phi$ is an angular
nuclear periodic coordinate which is constrained by some external agent,
as in (\cite{Moore}, \cite{Zwanziger}),
to change linearly in
time, namely $\phi = \omega t$ (thus, if $T$ is the time-period,
then $\omega = \frac{2\pi}{T}$),
$q$ is a radial coordinate,
$E_{el}$ is a constant and $G_i (q ,\phi)$; $i=1,2$ are two functions to be defined later.

The Schrodinger equation is ($\hbar=1$):
\beq
i \frac{\partial \Psi}{\partial t} = H \Psi.
\label{sch}
\enq
and this will be solved approximately to the first order in $\frac{q G}{E_{el}}$,
for the case that the ground state is an electronic doublet.
In a representation, adopted from \cite{Longuet}, this
doublet is described in terms of the electronic functions $\cos \theta$ and
$sin \theta $ and therefore $\psi$ can be expressed as: (\cite{Moore},\cite{Baer})
\beq
\Psi = \chi_1 (t) \cos \theta + \chi_2 (t) \sin \theta
\label{psichi}
\enq
In what follows \er{sch} will be solved for the initial conditions: $\chi_1( t=0) = 1$  and
$\chi_2 (t=0) = 0$.
Replacing $\chi_1 (t)$ and $\chi_2 (t)$ by $\psi_+ (t)$ and  $\psi_- (t)$ defined as:
\beq
\psi_{\pm} (t) = \frac{1}{2} \exp(i \frac{1}{2} E_{el} t)(\chi_1 \mp i \chi_2)
\label{psipm}
\enq
we get the corresponding equations for $\psi_{+} (t)$ and  $\psi_{-} (t)$:
\beq
i \dot {\psi}_{+} = -\frac{1}{2} \tilde{G} {\psi}_{-}
\qquad {\rm and} \qquad
i \dot {\psi}_{-} = -\frac{1}{2} \tilde{G}^* {\psi}_{+}
\label{eq2}
\enq
where $\tilde{G}$ is defined as: $\tilde{G} = G_{1} + i G_{2}$, and the dot represents
the time derivative.

Next we eliminate ${\psi}_{-}$ from \er{eq2} to obtain a single,
second order equation for $\psi_{+}$:
\beq
\ddot {\psi}_{+} - \dot{\ln (\tilde {G})} \dot {\psi}_{+} +
\frac{1}{4} |\tilde G|^2 {\psi}_{+} = 0.
\label{psipeq}
\enq
Writing $\tilde {G} = |\tilde {G}| \exp (i \Phi)$ we shall be interested in cases
where $|\tilde {G}|$ is constant, so that only $\Phi$ is time-dependent.
Thus \er{psipeq} becomes:
\beq
\ddot {\psi}_{+} - i \dot{\Phi} \dot {\psi}_{+} + \frac{1}{4} |\tilde {G}|^2 {\psi}_{+} = 0.
\label{psipeq2}
\enq
Once \er{psipeq2} is solved we can obtain $\chi_1 (t)$, the eigen-function for the initially
populated state. Usually, this is a fast oscillating function of t where the oscillations are
caused by the "dynamical phase" $\frac{1}{2} |\tilde {G}| t$.
This oscillatory component is eliminated upon multiplying  $\chi_1 (t)$ by $\exp(-\frac{1}{2} i |\tilde {G}| t)$.
In what follows we consider the smoother function
$\eta(t)$ defined as:
\beq
\eta(t) = \chi_1 (t) \exp(-\frac{1}{2} i |\tilde {G}| t)
\label{eta}
\enq
Our aim is the study of the time dependence of the phase $\gamma (t)$ defined through the
expression:
\beq
\eta(t) = \rho(t) \exp (i \gamma(t))
\label{etarhogam}
\enq
with $\rho(t)$ and $\gamma(t)$ real. Once  $\eta(t)$ is derived there
 are several ways to extract  $\gamma(t)$ we
shall use the following two: (a) the first is the following analytical
representation of the open path phase given by Pati \cite{Pati}:
\beq
\gamma(t)= \Im (\ln (\eta(t)))
\label{gam1}
\enq
where $\Im$ stands for the imaginary part of the expression in the parentheses.
Equation (\ref{gam1}) is
used for analytical purposes, as presented below.
Special emphasis will be put on $\gamma(t)$ at $t = T$ where $T$ is the period of the external
field. The case of an arbitrary $T$ will be discussed only briefly and we will be mainly
interested in the adiabatic case where $T$ is large, namely $T \gg |\tilde {G}|^{-1}$
for which $\gamma(t=T)$ becomes the topological (Berry) phase $\beta$.
In what follows we distinguish between two cases:
(a)  The degenerate case for which the functions in \er{ham} are given by:
\beq
G_1 (q,\phi) = G q \cos (\phi), \qquad G_2 (q,\phi) = G q \sin (\phi).
\label{Gidefdeg}
\enq
Where $G$ is constant.
We term it the degenerate case because the two lowest eigenvalues of \er{ham} become
equal in the $(q, \phi)$ plane at $q = 0$. It is also noticed that: $\tilde {G} = G q$.
 (b)  The non-degenerate case.
This is characterized by the condition that $G_1=0$ and $G_2=0$ cannot be simultaneously
satisfied for real $q$ and $\phi$.It is not trivial to achieve this by a simple change of the
expressions in \er{Gidefdeg}, since, e.g., adding a constant will only
displace the real root, as has
been previously discussed \cite{Yahalom}. However, non-degeneracy can be attained upon replacing
$G_2$ by a quadratic polynomial in $q \sin(\phi)$, such that the polynomial has
no real roots. A term
of this form is physically realizable in a low-symmetric molecular environment. (In a
realistic case of non crossing potential energy surfaces for NaFH, the expression
constructed for $G_2$ is very complicated \cite{Topaler}). Unfortunately, the
\er{psipeq} cannot be
solved analytically for a general polynomial $G_2$. Below (in section \ref{TNDC})
we present an approximate solution for a case
that a degeneracy is
encountered neither at $q = 0$ nor at any other real $q$-value.
This is achieved by the choice:
\beq
G_1 (q,\phi) = G q \cos (\phi), \qquad {\rm and} \qquad
G_2 (q,\phi) = \sqrt{(G q)^2 \sin^2 (\phi) + \mu^2}
\label{Gidef}
\enq
The quantity $\mu$ is related to the separation between the two potential energy surfaces.

If we expand the square
root for small $q$, we indeed recover a quadratic polynomial approximation,
that has no real roots.
It is noticed that now $|\tilde G| = \sqrt{G_1^2+G_2^2} = \sqrt{(G q)^2+ \mu^2}$.
In what follows it is assumed for
simplicity that the particle trajectory is on the circle $q = 1$.

\subsection{The Degenerate Case}
\label{TDC}

We start by considering the degenerate case and therefore in \er{psipeq}
$\Phi \equiv \phi = \omega t$ and $|\tilde G| = G$ as already mentioned,
(see \er{Gidefdeg}). As a result, \er{psipeq2} becomes:
\beq
\ddot {\psi}_{+} - i \omega \dot {\psi}_{+} + \frac{1}{4} G^2 {\psi}_{+} = 0.
\label{psipeq3}
\enq
The solution of this equation (as well as that of a similar equation for ${\psi}_{-} (t)$) can
be written in terms of trigonometric functions.
Returning to the original $\chi$-functions we get
for $\chi_1 (t)$ the following explicit expression:
\ber
{\chi_1} &=& \cos(k t) \cos(\frac{1}{2} \omega t) + \frac{\omega}{2 k}
\sin(k t) \sin(\frac{1}{2} \omega t) \nonumber \\
& + & i \frac{G}{2 k} \sin(k t) \cos(\frac{1}{2} \omega t)
\label{chi1}
\enr
where $k$, defined as:
\beq
k = \frac{1}{2} \sqrt{G^2 + \omega^2}
\label{kdef}
\enq
forms, together with $\omega$, two characteristic periodicities of the system.

\begin{figure}
\vspace{13cm}
\begin{picture}(1,1)
 \end{picture}
\includegraphics{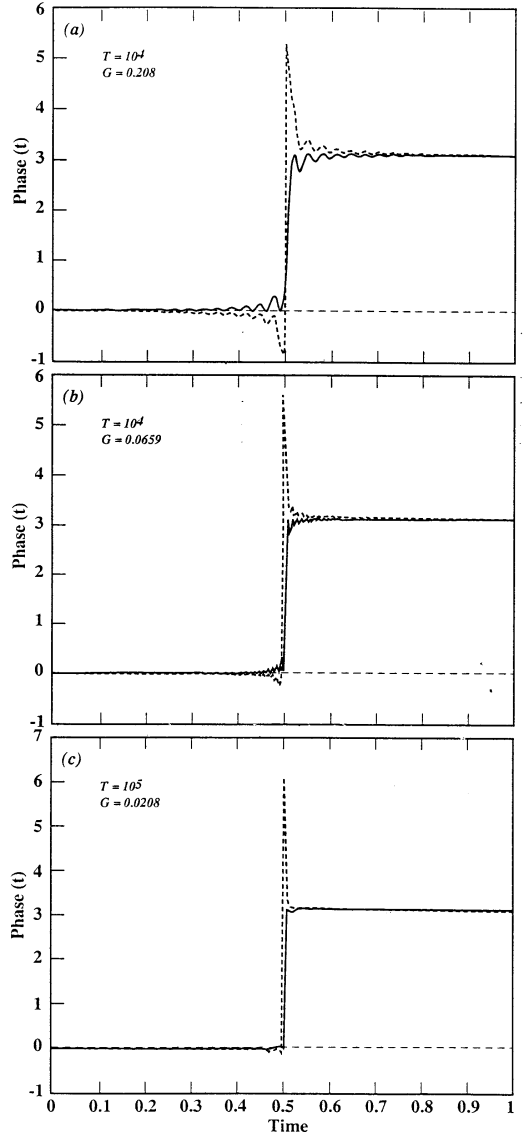}
\caption{The time-dependent phase factor $\gamma$ as a
function of time calculated for the degenerate case. The
calculations were done for different values of the external field
frequency \protect$\omega (=\protect\frac{2\pi}{T})\protect$, and
the coupling intensity \protect$ G \protect$. In all three cases
the Rabi oscillations are of a time period
\protect$(\frac{2\pi}{k})\protect$ where \protect$k$ is defined as
\protect$ k = \protect\frac{1}{2} \protect\sqrt{G^2 + \omega^2}
\protect$. In each sub-figure two curves are shown; one, drawn as
a full line, is the curve calculated employing equation
(\protect\ref{chi1}) and the other, drawn as a dashed line, is the
approximate curve calculated employing the Kramers-Kronig relation
shown in \eqs (\protect\ref{lamkk}-\protect\ref{lamkk3}) and
(\protect\ref{gam3}). (a) \protect$T=10^4$, \protect$G=0.01$. (b)
\protect$T=10^4$, \protect$G=0.02$. (c) \protect$T=10^5$,
\protect$G=0.01$.} \label{1}
\end{figure}

In Fig. \ref{1} are shown several $\gamma(t)$-functions as calculated
 for three different values of $G$
and $T (=\frac{2\pi}{\omega})$. It is noticed that as $T$ increases,
namely as the adiabatic limit is
approached, $\gamma(t)$ tends to a step function and $\beta$- the Berry phase-
 reaches the value of $\pi$.
This behavior is also derived analytically as follows.

Considering the case that $T \rightarrow \infty$ (or $\omega
\rightarrow  0$), one can show employing \eqs (\ref{gam1})
and (\ref{chi1}) for the adiabatic case, that $\gamma (t)$
takes the form (discarding second order terms in $\omega$):
\beq
\lim_{T \rightarrow \infty} (\gamma (t)) = \Im \{
\ln [\cos (\frac{1}{2} \omega t) + O (\omega)]\}
\label{gammchi}
\enq
Having this expression it is recognized that since
$\cos (\frac{1}{2} \omega t) > 0$ for $t \leq \frac{T}{2}$ and
$\cos (\frac{1}{2} \omega t) < 0$ for $t \geq \frac{T}{2}$
it follows that $ \gamma (t) \simeq 0$ for $0 \leq t \leq
\frac{T}{2}$ and $ \gamma (t) \simeq \pi $ for $\frac{T}{2}
\leq t \leq T$. This also implies that the topological
(Berry) phase $\beta \simeq \pi$. From Fig. \ref{1} it is noticed that,
when the adiabatic limit is approached (namely, $T \rightarrow \infty$),
$\gamma(t)$ becomes a step function.
The step takes place at $t \sim \frac{T}{2}$.
It is therefore of interest to study the behavior of $\gamma (t)$ in
the vicinity of $t = \frac{T}{2}$.
Thus expanding $\gamma (t)$ around this value and keeping only first order
terms in $(t - \frac{T}{2})$ yield:
\beq
\gamma (t \approx \frac{T}{2}) = \Im \{
\ln [\frac{T}{2} - t + \frac{1}{k} \sin kt \exp (ikt)]\}
\label{gammchi2}
\enq
It is noticed that around $t = \frac{T}{2}$ the phase factor
$\gamma (t)$ oscillates (Rabi oscillations) and its
periodicity is $(\frac{2\pi}{k})$.
These oscillations become more frequent the larger is the value of
the product $GT (\gg 1)$.

In order to obtain the phase using \er{lamkk3} we have to construct from
$\chi_1 (t)$, which when
analytically continued to the complex plane becomes $\chi_1 (z)$,
a new function that fulfills the
requirements imposed on $\tilde \psi (z) $.
The complex function $\chi_1 (z)$ is obtained by replacing in
\er{chi1}, the variable $t$ by $z$ defined as:
\beq
z = \tau \exp(i \theta); \qquad {\rm where} \qquad 0 \leq \theta \leq 2 \pi
\qquad {\rm and} \qquad \tau > 0.
\label{ztau}
\enq

The first requirement imposed on $\tilde \psi (z) $
is that it does not have zeros in the
lower half plane. The newly formed function $\chi_1 (z)$ has,
 in general, zeros
 in the lower half plane. But we have been able to show generally that
near the adiabatic limit there are no zeros for the ground state in the lower half plane.
[Moreover, a detailed numerical study showed that when the ratio of inverse periods
$(\frac{k}{\omega})=$ integer, the zeros (of the ground state) are located in
 the upper half plane (including the
real axis). For the near adiabatic situation where $(\frac{k}{\omega})$ is large,
 the requirement $\frac{k}{\omega}=$ (a
large) integer can (on physical grounds) differ only insignificantly from neighboring
values of $\frac{k}{\omega}$ that are non integral.
We thus have two independent reasons for the
assertion regarding the location of zeros in the near adiabatic case. This is also confirmed
by our graphical results in Figures \ref{1} and \ref{2},
which clearly show the increasing validity of
the integral relations, as the adiabatic limit is approached, upon going from (a) to (c), and
this even though the ratio $(\frac{k}{\omega})$ is not chosen to be an integer.]
The second requirement imposed on $\tilde \psi (z) $ is that it becomes equal to $1$ along the
infinite semi-circle on the lower half of the complex plane. From \er{chi1} it is readily
seen that for $\tau \rightarrow \infty$ the
function $\chi_1 (z)$ in the adiabatic limit becomes (for $ \theta > \pi$):
\beq
\lim_{\tau \rightarrow \infty} \chi_1 (z) =
\frac{1}{2} \exp (i (k+ \frac{1}{2} \omega) \tau e^{i \theta})
\label{limchi1}
\enq
 or
\beq
\lim_{\tau \rightarrow \infty} \chi_1 (z) =
\frac{1}{2} \exp (-(k+ \frac{1}{2} \omega) \tau \sin \theta)
\exp (i(k+ \frac{1}{2} \omega) \tau \cos \theta)
\label{limchi2}
\enq
Therefore multiplying $\chi_1 (z)$ by
$2 \exp (-i (k+ \frac{1}{2} \omega) \tau e^{i \theta})$
yields the function $\tilde \psi (z) $ which
becomes equal to $1$ along the infinite-semi circle. Thus the function to be employed in
\er{lamkk3} is $\tilde \psi (t)$  defined as:
\beq
\tilde \psi (t)=2 \chi_1 (t) \exp (-i (k+ \frac{1}{2} \omega) t)
\label{psidef}
\enq
Combining Eqs. \er{lamkk}, \er{lamkk3}, \er{eta}, \er{etarhogam} and \er{psidef}
we obtain the final expression for the phase
$\gamma (t)$ up to a linear function of time that follows from the KK equations:
\beq
\gamma (t)= \frac{2 t}{\pi} P \int_{0}^{\tilde T} d t' \ln(\tilde \Gamma(t'))
\sum_{n=0} \frac{1}{(t'+ N \tilde T)^2-t^2}
\label{gam3}
\enq
where $\tilde \Gamma(t)$ is is absolute value of $\chi_1 (t)$.
It is important to emphasize that $\tilde T$ is not
necessarily equal to $T$ (in our particular case $\tilde T$ is equal to $2T$).
In Fig. \ref{1} is presented $\gamma (t)$ also as calculated from \er{gam3}.
The results along the
interval $0 \leq t \leq (\frac{T}{2})$ were taken as they are but
those along the interval $\frac{T}{2} \leq t \leq T$ were
found to be $2 \pi$ below the values obtained by the direct method. We added to each of the
calculated values the physically unimportant magnitude $2 \pi$. The comparison between the
curves due to the two different calculations reveals a reasonable fit which improves when
either T or G become large enough, namely upon approaching the adiabatic limit. Even
the (Rabi) oscillations at the near adiabatic limit are well reproduced by the present
theory. Moreover the theory yields the correct geometrical phase. It is also important to
mention that when we are far from the adiabatic limit the fit is less satisfactory. However,
we also found that for the choices of $k$ which make the function $\chi_1 (t)$ periodic, namely
when $(\frac{k}{\omega})=$ integer,  the agreement resurfaces \cite{Baer}.

\section{The Non-Degenerate Case}
\label{TNDC}

This arises when $ \mu \neq 0$ (see \er{Gidef}).
As a result we obtain
for $|\tilde G|$, $\Phi$ and $\dot \Phi$ the following expressions:
\beq
|\tilde G| = \sqrt{G^2+\mu^2}; \qquad
\Phi= \arccos (p \cos \omega t); \qquad
\dot \Phi = p \omega \frac{ \sin \omega t}{\sqrt{1 - p^2 \cos^2 \omega t}}
\label{cphidphi}
\enq
where $p$ is defined as $p=\frac{G}{\sqrt{G^2+\mu^2}}$.
In what follows we consider only the case when
$\mu$ is small enough so that $|\tilde G|$ and $\Phi$
are, as before, equal to $G$ and $ \omega t$,
respectively, but $\dot \Phi$ will be written as:
$\dot \Phi = \omega \frac{ \sin \omega t}{|\sin \omega t|}$.
Thus \er{psipeq2} becomes:
\beq
\ddot {\psi}_{+} \mp i \omega \dot {\psi}_{+} + \frac{1}{4} |\tilde G|^2 {\psi}_{+} = 0.
\label{psipeqnd}
\enq
where the minus sign is for the $0 \leq t \leq (\frac{T}{2})$ -
the first half period and the plus sign for $\frac{T}{2} \leq t \leq T$
- the second half. For the first half we have the same equation as before and
therefore also the same solution (see \er{chi1}). As for the second half period we obtain a
somewhat more complicated expression for the solution due to the matching of the two
solutions at $t = \frac{T}{2}$. Thus:
\ber
\chi_1 (t) &=& e^{-i \pi} \{ \cos(k t) \cos(\frac{1}{2} \omega t) - \frac{\omega}{2 k}
\sin(k (T-t)) \sin(\frac{1}{2} \omega t) \nonumber \\
& + & i \frac{G}{2 k} \sin(k t) \cos(\frac{1}{2} \omega t) \}
\nonumber \\
& - & \frac{\omega}{2 k^2} \sin (\frac{k T}{2 }) \sin(k (t - \frac{1}{2} T))
[\omega \cos(\frac{1}{2} \omega t) + i G \sin(\frac{1}{2} \omega t) ]
\label{chi1nd}
\enr
In order to obtain the phase factor for the adiabatic case,\ \er{gam1} is applied as before,
where $\chi_1 (t)$ is given by \er{chi1nd}. We employed \er{chi1nd}
to calculate $\gamma (t)$ based on the KK dispersion relations.

\begin{figure}
\vspace{15cm}
\includegraphics{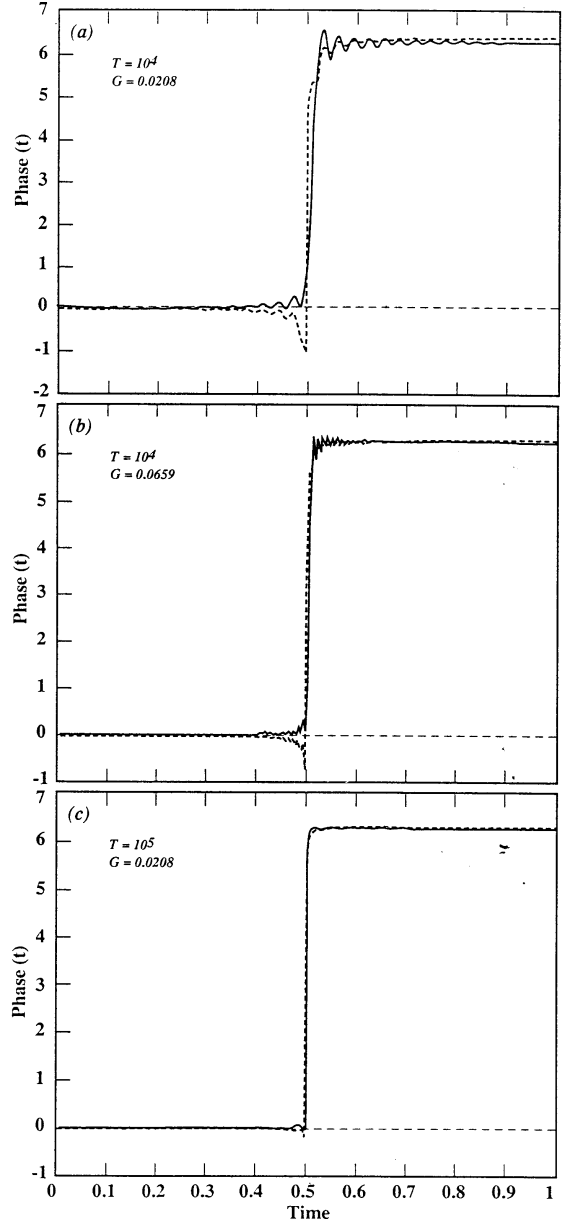}
\caption{The time dependent phase factor
\protect$\gamma$ as a function of time calculated for the
non-degenerate case. Details as in Fig.\protect\ref{1}. The
approximate curve was calculated employing \eqs
(\protect\ref{lamkk}-\protect\ref{lamkk3}) and
(\protect\ref{gam3}) for \protect$0 \leq t \leq (\frac{T}{2})$ and
\eqs (\protect\ref{lamkk}-\protect\ref{lamkk3}) and
(\protect\ref{chi1nd}) for \protect$\frac{T}{2} \leq t \leq T$.
(a) \protect$T=10^4$, \protect$G=0.01$. (b) \protect$T=10^4$,
\protect$G=0.02$. (c) \protect$T=10^5$, \protect$G=0.01$.}
\label{2}
\end{figure}

In Fig. \ref{2} are presented the results due to the two types of calculations as obtained for
three sets of values of the parameters $G$ and $T$. It is noticed that when either $T$ or $G$ become large
enough (namely, approaching the adiabatic limit), as in the previous case, a
reasonably good fit is obtained between the results due to the direct calculations and the
ones based on the KK relations (\eqs (\ref{lamkk3}) and (\ref{chi1nd})).
Moreover, in this case, too, this new formalism yields the correct geometrical phase.

The same analytic treatment can be done for the non-degenerate
two-state model.
Considering again the case that $T \rightarrow \infty$ (or $\omega
\rightarrow  0$), but for \er{chi1nd}, we obtain that $\gamma(t)$
takes the form:
\beq
\lim_{T \rightarrow \infty} (\gamma (t)) = \Im \{
\ln [\cos (\frac{1}{2} \omega t) + O (\omega)]\}
 + \Theta (t - \frac{T}{2}) \pi
\label{gammchind}
\enq
where $\Theta (x)$ is the Heavyside function
defined as being equal to zero for $x < 0$ and equal
to $1$ for $x > 0$.
It is noticed that the sign of the expression in the square
brackets is positive
for $0 \leq t \leq  \frac{T}{2}$ which means that
the phase factor is altogether zero (because also $\Theta (x)
=0$)
but the sign is positive for $\frac{T}{2}\leq t \leq T$
and therefore altogether $\gamma (t) = 2 \pi$
and this leads to a topological (Berry) angle $\beta = 2 \pi$.
This result is expected because the Berry phase has to
be zero (or $2 \pi$) in the case of no degeneracy.

\section{Conclusions}
\label{Conc}

The expression of the degeneracy and near-degeneracy dichotomy in the topological
phase is the main subject of this paper. The respective values of $\pi$
and $2 \pi$ after one revolution (seen in Figures \ref{1} and \ref{2}, respectively)
obtained in a two-stage model confirm the expectations. However, on the way
to this result we earned some new results and insights. Oscillations near the
half period stage were found (\er{gammchi2}) and explained. We also studied the
tendency of this and of other features in the "connection" (namely the non-cyclic
phase) with the approach to adiabatic (slow) behavior.

An attempt has been made in this article to establish a link between the time
dependent phase (and its particular value,
the topological phase, after a full revolution)
with the corresponding amplitude modulus.
To establish this relation we considered two alternative
two-state models, exposed to an external field,
under adiabatic and quasi-adiabatic
conditions (\cite{Aharonov}, \cite{Yahalom}).
The two types of models are physically different:
(a) one model contains an
(ordinary Jahn-Teller type) degeneracy at a point in configuration space;
(b) the second is
characterized by a nearly (in fact, non-)degenerate situation
(of the pseudo-Jahn-Teller type(\cite{Englman})) where the
two eigenvalues approach
each other at some point in configuration
space but do not touch.
In Figs. \ref{1} and \ref{2} are presented time dependent phases and the
(Berry) topological phases for these two models calculated in
two different ways: once directly by employing equation
(\ref{gam1})
and once by using the KK relations which led to \er{lamkk3}.
Essentially these findings suggest that one may be able to
obtain the time dependence
of the phase from a series of time-dependent measurements
of relative populations of a
given state.
We end by offering the following interpretation for our findings:
The phase on the left-hand side of \eqs (\ref{lamkk}-\ref{lamkk3})
is not a "physical observable" in the
conventional sense since no hermitian operator is associated
with it \cite{Wolf}. Yet, phases have
been observed in interference and other experiments \cite{Bitter}.
In the present formulation, \eqs.
(\ref{lamkk}-\ref{lamkk3}) associate the observable phase of the
wave function with $\Gamma(t)$ (the observable
probability amplitude) through integral expressions,
in a similar way to that done in  Ref.
\cite{Shapiro} for radiation fields.

\end{document}